# Early and Late Time Acoustic Measures for Underwater Seismic Airgun Signals In Long-Term Acoustic Data Sets

Peter. Dugan[1]
([Peter.Dugan@cornell.edu](Peter.Dugan@cornell.edu))

Melania Guerra[2], Dimitri Ponirakis[1], Holger Klinck[1] and Christopher W. Clark[1]

[(1)]Bioacoustics Research Program, Cornell Lab of Ornithology, Cornell University
Ithaca, NY 14850, USA

[(2)] Applied Physics Laboratory, University of Washington
Seattle, WA 98105, USA

*Abstract*—This work presents a new toolkit for describing the acoustic properties of the ocean environment before, during and after a sound event caused by an underwater seismic airgun. The toolkit uses existing sound measures, but uniquely applies these to capture the early time period (actual pulse) and late time period (reverberation and multiple arrivals). In total, 183 features are produced for each airgun sound. This toolkit was utilized on data retrieved from a field deployment encompassing five marine autonomous recording units (MARU's) during a 46-day seismic airgun survey in Baffin Bay, Greenland. Using this toolkit, a total of 147 million data points were identified from the Greenland deployment recordings. The feasibility of extracting a large number of features was then evaluated using two separate methods: a serial computer and a high performance system. Results indicate that data extraction performance took an estimated 216 hours for the serial system, and 18 hours for the high performance computer. This paper provides an analytical description of the new toolkit along with details for using it to identify relevant data.

*Seismic Airgun, Feature Extraction, Passive Acoustic Monitoring.*

I. INTRODUCTION

Receding sea ice in the Arctic is enabling an increased use of seismic technologies to explore uncharted polar areas for fossil fuels. Specialized equipment called seismic airgun arrays are commonly used to expeditiously locate pockets of fossil fuels beneath the ocean floor. Airguns are typically fired in a systematic way every 10- 20 s, generating periodic pulses with source levels which can exceed 260 $dB_{pp}$ re. 1 μPa [1]. The characteristics of the signal pulse may vary based on several factors [1], and the resulting high-energy acoustic signals can propagate long distances [2, 3]. Increasing noise levels in the ocean [4, 5] along with evidence that marine mammals are sensitive to sounds have raised concerns about the potential impacts of seismic exploration on the environment [1, 6-8]. The goal of this study was to develop a toolset which allows scientists to efficiently detect and extract a large suite of temporal and spectral properties (i.e. acoustic features) of airgun signals in long-term acoustic data sets. Application of digital M-weighted filters to model marine mammal hearing is briefly discussed. A series of three different weightings are recreated from [9] and used for this work.

The toolkit defines two different time intervals, the early time and late time. The first, is denominated "early time" and coincides with 90% of the energy of the direct pulse. The remainder of the acoustic energy following the direct pulse (reverberation and multiple arrivals) is referred to as the "late time." This late time interval lasts for several seconds and is further subdivided into shorter time windows of equal duration. The concept of dividing sounds into early and late time intervals is not new to the field of acoustics. Early work [10, 11] used a similar division to describe various psycho-acoustic properties as they relate to the listener; however no other research has been identified that uses these intervals to evaluate marine sounds produced from seismic airguns. Measurements were established for the *early time* and various *late time* windows. Each window includes four different combinations of measures which are sound pressure level, sound exposure level, sound equivalent level and the cumulative sound exposure level. The *early time* window has additional measures for describing the properties of the direct pulse. These include absolute early time, absolute late time, absolute time of positive peak pressure, positive peak pressure, absolute time of negative peak pressure, negative peak pressure, inter-pulse-interval (IPI), and late time intervals. The developed technique was tested on a data set spanning 5,520 channel hours of recordings. Data was collected using five Marine Acoustic Recording Units (MARUs) during a 46-day seismic airgun survey in Baffin Bay, Greenland.

Part one of this work discusses acoustic measures, which include various formulas that analytically describe properties of underwater sounds. Inner pulse interval (IPI) is presented as the impulse response to a series of two seismic airgun events, separated by a pulse interval. A proposed method for dividing impulse response into *early time* and *late time* intervals are defined in terms of the IPI, where various windows are adopted as a means to further divide the late time signal. Various sound pressure measures are defined, along with temporal characteristics of time varying signals. The context of the acoustic environment is further defined in terms of the potential listener, which is modeled using a series of filters specifically designed for marine mammal cetaceans.

Part two presents a compact table that summarizes the various features that are extracted from early and late time intervals by using the acoustic measures from part one. Results from Baffin Bay are presented, summarizing the number of features extracted as well as the time required to perform the data mining across the 46 day deployment. Paper is summarized by discussing results as well as the potential extensions to this work.

## II. ACOUSTIC MEASURES

### A. Early And Late Time Features

Figure 1 (upper panel) illustrates two seismic pulses received approximately 10 s apart. The lower panel of Figure 1 exemplifies a single seismic pulse, shown as the first box in the upper panel. Window segments are a common practice for describing portions of signal [12]. Description of the acoustic parameters is accomplished in two steps. First the seismic event is overlaid with a series of time windows, see Figure 1(upper panel), and two extraction routines are applied to the time series that is represented in each window. The window for the direct pulse is predetermined to 1.5 s, and its position is fixed relative to the signal peak values shown in Figure 1 (lower panel). Measurements from the 1.5 s window contain the main pulse as well as consecutive 1.0 s windows before the next pulse. Beginning and ending positions for the initial pulse are described by a defined process [13]. This process states that start and end times are defined as the integral of the 5th and 95th percentile energy values that happen before and after the peak amplitude of a pulse (1) (see Figure 1 [lower panel] for illustration).

$$\int_{5th}^{95th} p^2(t)dt . \quad (1)$$

The direct pulse contains peak pressure, negative peak pressure and peak-peak pressure measurements, as shown in Figure 1. According to Figure 1 (lower panel), the pulse is first located based on the peak pressure. Once the peak values are identified, the integral defined in (1) is used to determine start and stop times of the direct pulse. The time interval which coincides with (1) is given as $t(k)_{5th-95th}$, or shorthand by using brackets to represent the interval where $t_{[K]} = t(k)_{5th-95th}$. Various time based measures are computed including the time of occurrence for positive $t_{\bar{A}}$ and negative $t_{\bar{B}}$ peak pressure. Equations (2) and (3) represent linear (Pa) and log values (dB$_{pp}$ re 1 μPa) for positive and negative pressures respectively,

$$s(t_{\bar{A}}) = P_{\bar{A}}(\text{Pa}), \ P_{\bar{A}}(\text{dB}), \quad (2)$$

$$s(t_{\bar{B}}) = P_{\bar{B}}(\text{Pa}), \ P_{\bar{B}}(\text{dB}) . \quad (3)$$

This procedure is repeated for each direct pulse--a single airgun event. Details for this procedure are also spelled out in note 13 [13]. After the basic peak and time metrics are determined, a precise inter-pulse-interval (IPI) can be measured. The IPI is defined as the time between two successive pulses. Following from earlier discussions, the early time and late time are noted in (3) and (4) respectively,

$$t_{[K]} = t(k)_{5th-95th} \quad \text{early time}, \quad (4)$$

$$t_{[K][1..10]} = t_{[1..10]}(k) \quad \text{late time} . \quad (5)$$

The data extraction stage works on the *early time* and each *late time* window. For each window, measurements are calculated which include sound pressure level (SPL, in dB re 1 μPa), sound exposure level (SEL, dB re 1 μPa$^2$s) and sound equivalent level (L$_{EQ}$ in dB re 1 μPa$^2$). These are defined per equations noted at (6), (7) and (8) respectively:

$$SPL = 20\log_{10}\left(\frac{p_m(t)}{p_o}\right), \quad (6)$$

$$SEL = 10\log_{10}\int_0^T \left(\frac{p_m(t)}{p_o}\right)^2 dt, \quad (7)$$

$$L_{EQ} = 10\log_{10}\frac{1}{T}\int_0^T \left(\frac{p_m(t)}{p_o}\right)^2 dt . \quad (8)$$

In equations (6), (7) and (8), $p_m(t)$ is the measured acoustic pressure at time $t$ while $p_o$ represents the reference pressure (in water, 1 μPa) and $T$ represents the length of the window. As the 1 s resolution (T = 1 s) reduces equation (8) to equation (7), generating agreement between corresponding SEL and L$_{EQ}$ measurements, and thus some redundancy exists in the extracted features. Suggestions from several recent studies [9], intimate that cumulative Sound Exposure Level (CSEL) is recommended as an important metric for the assessment of impact of exposure on marine receptors (*e.g.*, marine mammals) to impulsive, anthropogenic sources. CSEL is theoretically defined as

$$CSEL = 10\log_{10}\left\{\frac{\sum_{n=1}^{N}\int_{0}^{T}p_{m}^{2}(t)dt}{p_{o}^{2}}\right\}. \quad (9)$$

CSEL can also be calculated as a collection of derived SEL pulses. This aggregate metric is useful for assessing the cumulative exposure of an animal to a sound field for an extended period of time.

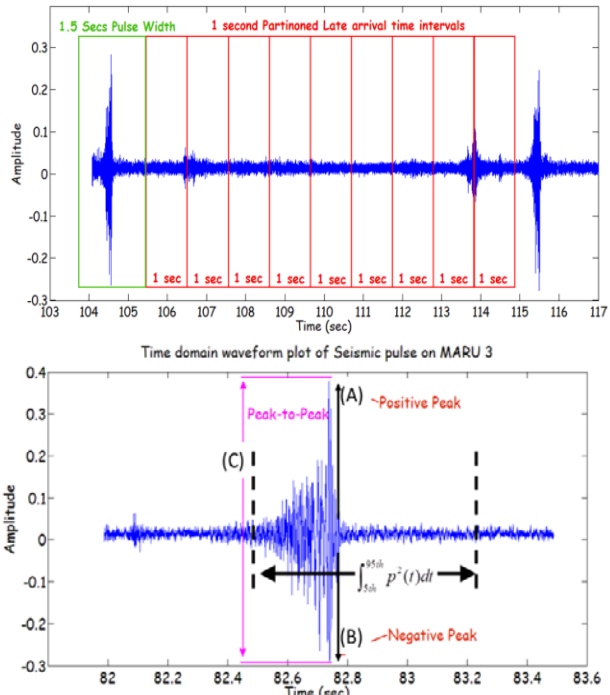

Figure 1,  Top: Two seismic events separated by inner pulse interval broken down into early time and late time intervals. Bottom:  Direct seismic air-gun pulse.

### B.  M-weighting filter

The acoustic environment is typically described from the perspective of a potential-listener.  Human hearing in air typically uses A,B and C weightings [11]. From the perspective of the potential listener, features extracted from early and late time intervals aim at quantitatively describing the acoustic environment. Unfortunately, aspects of marine mammal hearing are not completely understood [9, 13]. M-weightings, are measures based on pass band filters outlined in [9].  Simplified weightings selected for this work are reproduced from [15] and summarized in Table I.

Prior to extracting acoustic features, each seismic pulse is convolved with the corresponding M-weighting infinite impulse response (IIR) filter described in Table I. According to (Table I), $M_f$ is a series of three different weighting filters consisting of either Linear (flat frequency response), LFC (low frequency cetacean) or  MFC (mid-frequency cetacean).

### III.  METHODS

#### A.  Acoustic measures

Table II below summarize the features and associated time-pair values extracted for each registered seismic pulse event. A total of 11 different groups of feature combinations are shown.  Before any features were computed, three different weighting filters were used (Group 1), Table I, each generating a separate set of measures. Time based measures (Groups 2-9) were computed using equations (1). Groups 10-11 are SPL measures, computed using equations (6). Purposefully, the data have redundant information, such as the $L_{EQ}$ and SEL values for the 1-s windows case, which contain the same normalizing factor ($T = 1$ s) making these identical values. The rationale for doing this is to create a generic feature extraction algorithm that may  be applied to any time increment following the direct pulse event. This generic algorithm also facilitates cataloging the feature data for post-processing and follow-on research and analysis of the data products.

#### B.  Resulting Analytic Data

Table III summarizes the total number of points managed by the data-mining operation. A software routine computed measurements (Table II). Group 1 consists of 3 filters; Linear, LFC and MFC weightings.  Pair $t_{[K]}$ consists of the start and stop time values which satisfies the boundary conditions for $5^{th}$ and $95^{th}$ percentile using (1), this is shown as Group 2, Table II.  Remaining early time features are shown in groups 4-9, these coincide with (2-5). SPL features extracted from the early time window are shown in group 10, these coincide with (6-9).  Late time features consist of 10 features associated with start times for the late windows (Group 10) and 40 features associated with SPL values taken from ten late time windows (Group 11) totaling 50 late features.  Three weighted filters were used, one for each series of windows, resulting in 183 features per IPI interval, or 915 features when accounting for all 5 MARU's recorders.  Factoring in 160,122 pulses detected throughout the deployment results in 147 million feature points extracted.

TABLE I.  M-WEIGHTING FILTERS

| $M_f$ | $f_{lo}$ (Hz) | $f_{hi}$ (Hz) |
|---|---|---|
| Linear | Flat | Flat |
| LFC | 7 | 22,000 |
| MFC | 150 | 160,000 |

TABLE II. SEISMIC AIR GUN IPI MEASUREMENT SUMMARY, EARLY AND LATE ACOUSTIC FEATURES.

| Group | Number of Features | Reference | Window | Feature Description |
|---|---|---|---|---|
| 1 | 3 | $M_f$ | *Early, Late* | Weighting Filters *(Linear, LFC, MFC)*, convolved with early and late windows. |
| 2 | 1 | $t_{[K]}$ | *Early* | Early Time, $5^{th}$ and $95^{th}$ interval to estimate early pulse window |
| 3 | 10 | $t_{[K][1..10]}$ | *Late* | Reverberation start times for $1^{st}$ – $10^{th}$ late time window. |
| 4 | 1 | $t_{\bar{A}}$ | *Early* | Time of positive peak pressure, value from initial positive amplitude. |
| 5 | 1 | $P_{\bar{A}}(Pa)$ | *Early* | Positive peak pressure (Pa). |
| 6 | 1 | $P_{\bar{A}}(dB)$ | *Early* | Positive peak pressure (dB re 1 µP). |
| 7 | 1 | $t_{\bar{B}}$ | *Early* | Time of negative peak pressure (s). |
| 8 | 1 | $P_B(Pa)$ | *Early* | Negative peak pressure (Pa). |
| 9 | 1 | $P_{\bar{B}}(dB)$ | *Early* | Negative peak pressure (dB re 1 µP). |
| 10 | 4 | SPL, SEL, $L_{EQ}$, CSEL | *Early* | Direct pulse SPL measures for early window. (see equations 6-9) |
| 11 | 40 | SPL, SEL, $L_{EQ}$, CSEL | *Late* | Reverberation SPL measures for late windows [$1^{st}$, $2^{nd}$..$10^{th}$]. (see equations 6-9) |

*C. Data and Processing*

A total of 5 MARUs were deployed and continuously operated in Baffin Bay, Greenland for 46 days in 2010. The combined deployment consisted of 230 channel days (5,520 hours) of sounds. The soundscape was sampled at 16 kHz with a 12 bit sample depth. The total data set comprised 587 GB of data. The feasibility of automatically extracting high resolution features was tested using two different experimental methods. Both methods used a Dell Power Edge 610T multi-core server, equipped with dual octal cores of 2.67 GHz, Intel Xeon X650 processor totaling 16 nodes. The only difference between the experiments was the number of cores used for data mining of the acoustic features. The first experiment, *Serial Method*, used a single core only. The second experiment, or *Parallel Method*, utilized 12 processing cores. This approach required a special software package designed for high performance computing applications [16]. The HPC software provided the capability to distribute the processing across each node without requiring any modification to the data mining routines. Therefore the feature extraction code was identical for both methods.

TABLE III. NUMBER OF EXTRACTED FEATURES

| A | Number of weighting filters (signal streams) | 3 |
|---|---|---|
| B | Early Time Features ( feature points) | 11 |
| C | Late Time Features, (feature points) | 50 |
| D | Number of recording Units (Number of MARU's) | 5 |
| E | Pulses detected (Number of pulses) | 160,122 |
|  | Total number of data points extracted (A)(B+C)(D)(E) | ~147x$E^6$ feature points |

IV. RESULTS

*A. Extraction Points*

In accordance with the acoustic features outlined in Table II, this data mining effort for Baffin Bay produced 147 million data points (Table III). According to Table III, M-weighting pre-processing filters provided three independent sets of early and late IPI measures for Linear, LFC and MFC filters. Early and late windows were applied to each pulse, features extracted based on Table II. Early time windows required 11 features and late time windows 50 features, totaling 61 features for each seismic pulse. Applying three M-weighting filter across 5 MARU's, the entire deployment produced 147 million features.

*B. Serial and Parallel Processing Methods*

Based on the total measures in Table III, the HPC machine was used to stage and run 5,520 channel hours of data. Results are summarized in Table IV. Upon inspection of the performance of the *Serial Method*, it was decided that the process was too slow for analysis applications. A full runtime metric was estimated by processing only 1 of the 5 channels. Channel 3 was analyzed for the full 46 days, completing in 43 hours. Since each channel contained the same amount of data, but acoustically represented different GPS deployment locations a total runtime for the serial case was estimated by scaling by a factor of five times the single channel run, Table IV. *Parallel Method* was run using the same software routines as used in the serial case. Twelve computer cores were used to process the recordings as described in [16] and [17]. To make equal comparisons between the methods, channel 3 was processed by itself, with a total runtime of 3.8 hours. The complete dataset was then staged and run using 12 cores with a total runtime of 18.1 hours. Table IV summarizes these values to the nearest hour.

TABLE IV. RUNTIME PERFORMANCE, NEAREST HOUR

|  | Runtime (Ch. 3) | Runtime (Ch. 1-5) |
|---|---|---|
| Serial Method | 43 hours | 216 hours (estimated) |
| Parallel Method | 4 hours | 18 hours |

## V. DISCUSSION

The analysis described herein accomplished two significant goals. The first goal was to develop a toolset for extracting early and late acoustic measures from seismic pulses. In all, 160,122 pulses were automatically mined using early and late time interval measures to describe the sound pressure levels for various acoustic metrics. The data mining operation produced over 120 million measures for future analysis. The second goal was to utilized a high-speed, high performance computer system to process a complete survey effort for seismic events with a field deployment lasting 46 days, using five MARU sensors yielding 5,520 channel hours of data. Using serial processing, 147 million feature points would require roughly 216 hours, or 9 days, of execution time; making this approach somewhat impractical for studies of this scale. The advanced HPC system was able to use 12 processors that required 18 hours of runtime; saving of 198 hours or 8.25 days making this a practical solution for future work.

## VI. ACKNOWLEDGMENT

This work was partially funded by Cairn Energy, LLC, 50 Lothian Road, Edinburgh EH3 9BY. HPC funding was provided by the National Oceanic Partnership Program (NOPP); the Marine Mammals and Biology Program at the Office of Naval Research (ONR), award number N000141210585; and the National Fish and Wildlife Foundation (NFWF) award number 0309.07.28515.